\documentstyle[12pt,epsfig]{article} 

\expandafter\ifx\csname mathrm\endcsname\relax\def\mathrm#1{{\rm #1}}\fi

\makeatletter
\if@twoside \oddsidemargin -17pt \evensidemargin 00pt
\else \oddsidemargin 00pt \evensidemargin 00pt
\fi
\expandafter\ifx\csname mathrm\endcsname\relax\def\mathrm#1{{\rm #1}}\fi

\makeatother

\def\beq{\begin{equation}}
\def\eeq{\end{equation}}
\def\beqar{\begin{eqnarray}}
\def\eeqar{\end{eqnarray}}
\def\barr#1{\begin{array}{#1}}
\def\earr{\end{array}}
\def\bfi{\begin{figure}}
\def\efi{\end{figure}}
\def\btab{\begin{table}}
\def\etab{\end{table}}
\def\bce{\begin{center}}
\def\ece{\end{center}}
\def\nn{\nonumber}

\def\text{\textstyle}

\def\al{\alpha}

\def\ga{\gamma}
\def\de{\delta}

\def\De{\Delta}

\def\refeq#1{\mbox{(\ref{#1})}}
\def\reffi#1{\mbox{Fig.~\ref{#1}}}

\def\refta#1{\mbox{Tab.~\ref{#1}}}
\def\citere#1{\mbox{Ref.~\cite{#1}}}
\def\citeres#1{\mbox{Refs.~\cite{#1}}}

\newcommand{\GeV}{\unskip\,{\mathrm GeV}}
\newcommand{\MeV}{\unskip\,{\mathrm MeV}}
\newcommand{\TeV}{\unskip\,{\mathrm TeV}}

\def\mathswitchr#1{\relax\ifmmode{\mathrm{#1}}\else$\mathrm{#1}$\fi}

\newcommand{\PW}{\mathswitchr W}
\newcommand{\PZ}{\mathswitchr Z}

\newcommand{\PH}{\mathswitchr H}

\newcommand{\Pb}{\mathswitchr b}

\newcommand{\Pt}{\mathswitchr t}

\def\mathswitch#1{\relax\ifmmode#1\else$#1$\fi}

\newcommand{\MW}{\mathswitch {M_\PW}}

\newcommand{\MZ}{\mathswitch {M_\PZ}}
\newcommand{\MH}{\mathswitch {M_\PH}}

\newcommand{\Mb}{\mathswitch {m_\Pb}}
\newcommand{\Mt}{\mathswitch {m_\Pt}}

\newcommand{\scrs}{\scriptscriptstyle}
\newcommand{\sw}{\mathswitch {s_{\scrs\PW}}}
\newcommand{\swtwo}{\mathswitch {s_{{\scrs\PW}, (2)}}}
\newcommand{\swone}{\mathswitch {s_{{\scrs\PW}, (1)}}}
\newcommand{\swtone}{\mathswitch {s_{{\scrs\PW}, (1), \Pt}}}
\newcommand{\swHone}{\mathswitch {s_{{\scrs\PW}, (1), \PH}}}
\newcommand{\cw}{\mathswitch {c_{\scrs\PW}}}

\newcommand{\GF}{\mathswitch {G_\mu}}

\newcommand{\alpz}{\alpha(\MZ^2)}
\newcommand{\alps}{\alpha_{\mathrm s}}

\newcommand{\ses}{self-en\-er\-gies}
\newcommand{\fea}{{\em FeynArts}}

\newcommand{\two}{{\em TwoCalc}}

\renewcommand{\Re}{\mathop{\mathrm{Re}}}

\def\draftdate{\relax}
\def\mda{\relax}
\def\mua{\relax}
\def\mla{\relax}
\def\draft{
\def\thtystars{******************************}
\def\sixtystars{\thtystars\thtystars}
\typeout{}
\typeout{\sixtystars**}
\typeout{* Draft mode!
         For final version remove \protect\draft\space in source file *}
\typeout{\sixtystars**}
\typeout{}
\def\draftdate{\today}
\def\mua{\marginpar[\boldmath\hfil$\uparrow$]%
                   {\boldmath$\uparrow$\hfil}%
                    \typeout{marginpar: $\uparrow$}\ignorespaces}
\def\mda{\marginpar[\boldmath\hfil$\downarrow$]%
                   {\boldmath$\downarrow$\hfil}%
                    \typeout{marginpar: $\downarrow$}\ignorespaces}
\def\mla{\marginpar[\boldmath\hfil$\rightarrow$]%
                   {\boldmath$\leftarrow $\hfil}%
                    \typeout{marginpar: $\leftrightarrow$}\ignorespaces}
\def\Mua{\marginpar[\boldmath\hfil$\Uparrow$]%
                   {\boldmath$\Uparrow$\hfil}%
                    \typeout{marginpar: $\Uparrow$}\ignorespaces}
\def\Mda{\marginpar[\boldmath\hfil$\Downarrow$]%
                   {\boldmath$\Downarrow$\hfil}%
                    \typeout{marginpar: $\Downarrow$}\ignorespaces}
\def\Mla{\marginpar[\boldmath\hfil$\Rightarrow$]%
                   {\boldmath$\Leftarrow $\hfil}%
                    \typeout{marginpar: $\Leftrightarrow$}\ignorespaces}
\overfullrule 5pt
\oddsidemargin -15mm
\marginparwidth 29mm
}

\begin{document}
\thispagestyle{empty}
\null
\hfill KA-TP-05-1997\\
\null
\hfill WUE-ITP-97.005\\
\null
\hfill hep-ph/9707510\\
\vskip .8cm
\begin{center}
{\Large \bf Higgs-mass dependence of two-loop\\[.5em]
corrections to $\De r$}
\vskip 2.5em
{\large
{\sc Stefan Bauberger}%
\footnote{Work supported by the German Federal Ministry for
Research and Technology (BMBF) under contract number
05 7WZ91P (0).}
}\\[1ex]
{\normalsize \it
Institut f\"ur Theoretische Physik, Universit\"at W\"urzburg, Am
Hubland, D-97074~W\"urzburg, Germany}\\[2ex]
{\large
{\sc Georg Weiglein}\\[1ex]
{\normalsize \it Institut f\"ur Theoretische Physik, Universit\"at
Karlsruhe,\\
D--76128 Karlsruhe, Germany}
}
\vskip 2em
\end{center} \par
\vskip 1.2cm
\vfil
{\bf Abstract} \par
The Higgs-mass dependence of the Standard Model contributions to the 
correlation between the gauge-boson masses is studied at the two-loop
level. Exact results are given for the Higgs-dependent two-loop
corrections associated with the fermions, i.e.\ no expansion in the
top-quark and the Higgs-boson mass is made. The results for the
top quark are compared with results of an expansion up to next-to-leading
order in the top-quark mass. Agreement is found within $30 \%$ of the
two-loop result.
The remaining theoretical uncertainties in the Higgs-mass
dependence of $\De r$ are discussed.
\par
\vskip 1cm
\null
\setcounter{page}{0}
\clearpage

The remarkable accuracy of the electroweak precision data allows to
thoroughly test the predictions of the 
electroweak Standard Model (SM) at its quantum level, where all the
parameters of the model enter the theoretical predictions.
In this way it has been possible to predict the value of the 
top-quark mass, $\Mt$, within the SM 
prior to its actual experimental discovery~\cite{mtexp},
and the predicted value turned out to be in impressive agreement with
the experimental result.

After the discovery of the top quark the Higgs boson remains the only
missing ingredient of the minimal SM. At the moment, the mass of the
Higgs boson, $\MH$, can still only rather mildly be constrained by
confronting the SM with precision data~\cite{datasum97}.
An important goal for the future is therefore a further
reduction of the experimental and theoretical errors, not only 
in order to obtain stronger bounds for the Higgs-boson mass but also 
to achieve improved sensitivity to effects of physics beyond the SM.

Concerning the reduction of the theoretical error due to missing 
higher-order corrections, in particular a precise prediction for the 
basic relation between the masses $\MW$, $\MZ$ of the vector bosons, the
Fermi constant $\GF$, and the fine structure constant $\al$ 
is of interest. This relation is commonly expressed in terms of the
quantity $\De r$~\cite{sirlin} derived from muon decay.
With the prospect of the improving accuracy of the measurement of the
W-boson mass at LEP2 and the Tevatron, the importance of $\De r$ for
testing the electroweak theory becomes even more pronounced.

At the one-loop level the largest contributions to $\Delta r$ in the SM
are the QED induced shift in the fine structure constant, $\De \al$, and
the contribution of the top/bottom weak isospin doublet, which gives
rise to a term that grows as $\Mt^2$.
The SM one-loop result for $\De r$~\cite{sirlin} has been supplemented by
resummations of certain one-loop contributions~\cite{resum,sirresum}. 
While QCD corrections at ${\cal O}(\al \alps)$~\cite{R7a,qcd2} and 
${\cal O}(\al \alps^2)$~\cite{qcd3} are available and the remaining
theoretical uncertainty of the QCD corrections has been estimated to be
small~\cite{YeRep,PGRing}, 
the electroweak results at the two-loop level have so far
been restricted to expansions in either $\Mt$ or $\MH$. The leading top-quark
and Higgs-boson contributions have been evaluated in
\citeres{vdBH,vdBV}. The full Higgs-boson dependence of
the leading $\GF^2 \Mt^4$ contribution was calculated in 
\citere{barb2}, and recently also
the next-to-leading top-quark contributions of 
${\cal O}(\GF^2 \Mt^2 \MZ^2)$ were derived~\cite{gamb}.

In the global SM fits to all available data~\cite{datasum97}, 
where the ${\cal O}(\GF^2 \Mt^2 \MZ^2)$ correction obtained in \citere{gamb} 
is not yet included, the error due to missing higher-order corrections 
has a strong effect on the resulting value of $\MH$, shifting the upper
bound for $\MH$ at 95\% C.L.\ by $\sim +100 \GeV$. 
For the different precision observables this error
is comparable to the 
error caused by the parametric uncertainty related to the experimental error
of the hadronic contribution to $\alpz$~\cite{hollwarsaw}.
In \citeres{PGRing,DGS} it is argued that inclusion of the 
${\cal O}(\GF^2 \Mt^2 \MZ^2)$ contribution will lead
to a significant reduction of the error from the missing higher-order
corrections.

Both the Higgs-mass dependence of the leading $\Mt^4$ contribution
and the inclusion of the next-to-leading term in the $\Mt$~expansion
turned out to yield important corrections. In order to further settle
the issue of theoretical uncertainty due to missing higher-order
corrections therefore a more complete calculation 
would be desirable, where no expansion in $\Mt$ or $\MH$ is made.

In this paper the Higgs-mass dependence of the two-loop contributions
to $\De r$ in the SM is studied. The corrections associated with the
fermions are evaluated exactly, i.e.\ without an expansion in the masses. 
The results are compared with the expansion up to next-to-leading order
in the top-quark mass.
\bigskip

The relation between the vector-boson masses in terms of the Fermi
constant reads~\cite{sirlin}
\beq
\MW^2 \left(1 - \frac{\MW^2}{\MZ^2}\right) = 
\frac{\pi \al}{\sqrt{2} \GF} \left(1 + \De r\right),
\eeq
where the radiative corrections are contained in the quantity $\De r$.
In the context of this paper we treat $\De r$ without resummations,
i.e.\ as being fully expanded up to two-loop order,
\beq
\De r = \De r_{(1)} + \De r_{(2)} + {\cal O}(\al^3) .
\eeq
The theoretical predictions for $\De r$ are obtained by calculating
radiative corrections to muon decay. 

{}From a technical point of view the calculation of top-quark
and Higgs-boson
contributions to $\Delta r$ and other processes with light external
fermions at low energies requires in particular the evaluation of
two-loop self-energies on-shell, 
i.e.\ at non-zero external momentum, while vertex and
box contributions can mostly be reduced to vacuum integrals. The
problems encountered in such a calculation are due to the large number
of contributing Feynman diagrams, their complicated tensor structure, 
the fact that scalar two-loop integrals are in general not expressible
in terms of polylogarithmic functions~\cite{ScharfDipl}, and due to the
need for a two-loop renormalization, which has not yet been worked out
in full detail. 

The methods that we use for carrying out such a calculation have been
outlined in \citere{sbaugw1}. The generation of the diagrams and
counterterm contributions is done with the help of the computer-algebra
program \fea\ \cite{fea}. Making use of two-loop tensor-integral 
decompositions, the generated amplitudes are reduced to a minimal set 
of standard scalar integrals with the program \two~\cite{two}. The
renormalization is performed within the complete on-shell 
scheme~\cite{onshell}, i.e.\
physical parameters are used throughout. The two-loop scalar integrals
are evaluated numerically with one-dimensional integral 
representations~\cite{intnum}. These allow a very fast calculation of the 
integrals with high precision without any approximation in the masses.
As a check of our calculations, all results presented in this paper have
been derived using a general $R_{\xi}$~gauge, which is specified by one
gauge parameter $\xi_i, \; i = \ga, Z, W,$ for each vector boson, and 
we have explicitly verified that the gauge parameters actually drop
out.
As input parameters for our numerical analysis we use unless otherwise
stated the following values: 
$\alpz^{-1} \equiv (1 - \De\alpha)/\alpha(0) = 128.896$,
$\GF = 1.16639\, 10^{-5} \, \GeV^{-2}$,
$\MZ = 91.1863\, \GeV$,
$\Mt = 175.6\, \GeV$,
$\Mb = 4.7\, \GeV$.

In order to study the Higgs-mass dependence of the two-loop
contributions to $\De r$ we consider the
subtracted quantity
\beq
\label{eq:DeltaRsubtr}
\De r_{(2), {\mathrm subtr}}(\MH) =
\De r_{(2)}(\MH) - \De r_{(2)}(\MH = 65\GeV),
\eeq
where $\De r_{(2)}(\MH)$ denotes the two-loop contribution to
$\De r$.
\smallskip

Potentially large $\MH$-dependent contributions are the corrections
associated with the top quark, since the Yukawa coupling of the Higgs 
to the top quark is proportional to $\Mt$, and the contributions which are 
proportional to $\De\al$. We first consider the
Higgs-mass dependence of the two-loop top-quark contributions
and calculate the quantity $\De r^{\mathrm top}_{(2), {\mathrm
subtr}}(\MH)$ which denotes the contribution of the top/bottom doublet
to $\De r_{(2), {\mathrm subtr}}(\MH)$.

{}From the one-particle irreducible diagrams obviously 
those graphs contribute
to $\De r^{\mathrm top}_{(2), {\mathrm subtr}}(\MH)$ that contain both
the top and/or bottom quark and the Higgs boson. 
It is easy to see that only two-point
functions enter in this case, since all graphs where the Higgs boson
couples to the muon or the electron may safely be neglected. 
Although no two-loop three-point function enters, there is 
nevertheless a contribution from the two-loop and one-loop vertex
counterterms. If the field renormalization constants of the W~boson
are included (which cancel in the complete result), the vertex
counterterms are separately finite. 

The technically most complicated contributions arise from
the mass and mixing-angle renormalization. Since it is performed in
the on-shell scheme, the evaluation of the W- and Z-boson self-energies
are required at non-zero momentum
transfer.%
\footnote{It should be noted at this point that in the context of this
paper the question is immaterial whether the
mass definition of unstable particles at the two-loop level should be 
based on the real part of the complex pole of the S~matrix or on the
real pole. For the contributions investigated here both definitions
are equivalent.}

Expressed in terms of the one-loop and two-loop contributions to 
the transverse part of the W-boson self-energy 
$\Sigma^{\PW}(p^2)$ and the counterterm
$\de Z^{\mathrm vert}$ to the $W^- \bar e \nu_e$ vertex, the quantity 
$\De r^{\mathrm top}_{(2), {\mathrm subtr}}(\MH)$ reads
\beqar
\lefteqn{
\De r^{\mathrm top}_{(2), {\mathrm subtr}}(\MH) = 
\biggl[ \frac{\Sigma^{\PW}_{(2)}(0) - 
\Re \Sigma^{\PW}_{(2)}(\MW^2)}{\MW^2} 
+ 2 \de Z^{\mathrm vert}_{(2)}  \nn } \\
&& {} + 2 \frac{\left(\Sigma^{\PW}_{(1), \Pt}(0) -
\Re \Sigma^{\PW}_{(1), \Pt}(\MW^2)\right) 
\left(\Sigma^{\PW}_{(1), \PH}(0) -
\Re \Sigma^{\PW}_{(1), \PH}(\MW^2)\right)}{\MW^4} \nn \\
&& {} + 2 \frac{\left(\Sigma^{\PW}_{(1), \Pt}(0) - 
\Re \Sigma^{\PW}_{(1), \Pt}(\MW^2)\right) 
\de Z^{\mathrm vert}_{(1), \PH}}{\MW^2}  \nn \\
&& {} + 2 \frac{\left(\Sigma^{\PW}_{(1), \PH}(0) - 
\Re \Sigma^{\PW}_{(1), \PH}(\MW^2)\right)
\de Z^{\mathrm vert}_{(1), \Pt}}{\MW^2} 
+ 2 \de Z^{\mathrm vert}_{(1), \Pt}
\de Z^{\mathrm vert}_{(1), \PH}
\biggr]_{\mathrm subtr}, 
\label{eq:Deltrtopsubtr}
\eeqar
where it is understood that the two-loop contributions to the \ses\
contain the subloop renormalization. The two-loop terms denote those
graphs that contain both the top quark and the Higgs boson, while for the 
one-loop terms the top-quark and the Higgs-boson contributions are
indicated by a subscript. The two-loop vertex counterterm is expressible
in terms of the charge counterterm $\de Z_e$ and the mixing-angle
counterterm $\de \sw/\sw$,
\beq
\de Z^{\mathrm vert}_{(2)} = \de Z_{e, (2)} - \frac {\de \swtwo}{\sw}
+ 2 \frac {\de \swtone}{\sw} \frac {\de \swHone}{\sw} 
- \de Z_{e, (1), \Pt} \frac {\de \swHone}{\sw},
\label{eq:dZvert2}
\eeq
and similarly the one-loop vertex counterterm is given by
\beq
\de Z^{\mathrm vert}_{(1)} = \de Z_{e, (1)} - \frac {\de \swone}{\sw},
\label{eq:dZvert1}
\eeq
with $\sw^2 \equiv 1 - \cw^2 = 1 - \MW^2/\MZ^2$.
For the Higgs-dependent fermionic
contributions the charge counterterm is related to
the photon vacuum polarization according to 
\beq
\de Z_{e, (2)} = - \frac{1}{2} \de Z_{AA, (2)} = 
\frac{1}{2} \Pi^{AA}_{(2)}(0) ,
\label{eq:dZe2}
\eeq
which is familiar from QED. The validity of \refeq{eq:dZe2} can also
be understood by observing that the contributions considered here are
precisely the same as the ones obtained within the framework of the 
background-field method~\cite{bfmlong}.

The mixing angle counterterm $\de \swtwo/\sw$ is expressible 
in terms of the on-shell two-loop W-boson and Z-boson \ses\ and additional
one-loop contributions,
\beqar
\lefteqn{
\frac{\de \swtwo}{\sw} = - \frac{1}{2} \frac{\cw^2}{\sw^2}
\left(\frac{\Re \Sigma^{\PW}_{(2)}(\MW^2)}{\MW^2} - 
\frac{\Re \Sigma^{\PZ\PZ}_{(2)}(\MZ^2)}{\MZ^2}\right) \nn} \\
&& {} 
- \frac{\de \swtone}{\sw} 
  \frac{\Re \Sigma^{\PZ\PZ}_{(1), \PH}(\MZ^2)}{\MZ^2} 
- \frac{\de \swHone}{\sw} 
  \frac{\Re \Sigma^{\PZ\PZ}_{(1), \Pt}(\MZ^2)}{\MZ^2}
- \frac{\de \swtone}{\sw} \frac{\de \swHone}{\sw} ,
\eeqar
where 
\beq
\frac{\de \swone}{\sw} = - \frac{1}{2} \frac{\cw^2}{\sw^2}
\left(\frac{\Re \Sigma^{\PW}_{(1)}(\MW^2)}{\MW^2} -
\frac{\Re \Sigma^{\PZ\PZ}_{(1)}(\MZ^2)}{\MZ^2}\right) .
\eeq
In \refeq{eq:Deltrtopsubtr}--\refeq{eq:dZvert1}
the field renormalization constants of the W~boson have been omitted.
In our calculation of $\De r^{\mathrm top}_{(2), {\mathrm subtr}}(\MH)$ 
we have explicitly kept the field renormalization constants of 
all internal fields and have checked
that they actually cancel in the final result. 

\begin{figure}[htb]
\bce
\setlength{\unitlength}{0.1bp}
\special{!
/gnudict 40 dict def
gnudict begin
/Color false def
/Solid false def
/gnulinewidth 5.000 def
/vshift -33 def
/dl {10 mul} def
/hpt 31.5 def
/vpt 31.5 def
/M {moveto} bind def
/L {lineto} bind def
/R {rmoveto} bind def
/V {rlineto} bind def
/vpt2 vpt 2 mul def
/hpt2 hpt 2 mul def
/Lshow { currentpoint stroke M
  0 vshift R show } def
/Rshow { currentpoint stroke M
  dup stringwidth pop neg vshift R show } def
/Cshow { currentpoint stroke M
  dup stringwidth pop -2 div vshift R show } def
/DL { Color {setrgbcolor Solid {pop []} if 0 setdash }
 {pop pop pop Solid {pop []} if 0 setdash} ifelse } def
/BL { stroke gnulinewidth 2 mul setlinewidth } def
/AL { stroke gnulinewidth 2 div setlinewidth } def
/PL { stroke gnulinewidth setlinewidth } def
/LTb { BL [] 0 0 0 DL } def
/LTa { AL [1 dl 2 dl] 0 setdash 0 0 0 setrgbcolor } def
/LT0 { PL [] 0 1 0 DL } def
/LT1 { PL [4 dl 2 dl] 0 0 1 DL } def
/LT2 { PL [2 dl 3 dl] 1 0 0 DL } def
/LT3 { PL [1 dl 1.5 dl] 1 0 1 DL } def
/LT4 { PL [5 dl 2 dl 1 dl 2 dl] 0 1 1 DL } def
/LT5 { PL [4 dl 3 dl 1 dl 3 dl] 1 1 0 DL } def
/LT6 { PL [2 dl 2 dl 2 dl 4 dl] 0 0 0 DL } def
/LT7 { PL [2 dl 2 dl 2 dl 2 dl 2 dl 4 dl] 1 0.3 0 DL } def
/LT8 { PL [2 dl 2 dl 2 dl 2 dl 2 dl 2 dl 2 dl 4 dl] 0.5 0.5 0.5 DL } def
/P { stroke [] 0 setdash
  currentlinewidth 2 div sub M
  0 currentlinewidth V stroke } def
/D { stroke [] 0 setdash 2 copy vpt add M
  hpt neg vpt neg V hpt vpt neg V
  hpt vpt V hpt neg vpt V closepath stroke
  P } def
/A { stroke [] 0 setdash vpt sub M 0 vpt2 V
  currentpoint stroke M
  hpt neg vpt neg R hpt2 0 V stroke
  } def
/B { stroke [] 0 setdash 2 copy exch hpt sub exch vpt add M
  0 vpt2 neg V hpt2 0 V 0 vpt2 V
  hpt2 neg 0 V closepath stroke
  P } def
/C { stroke [] 0 setdash exch hpt sub exch vpt add M
  hpt2 vpt2 neg V currentpoint stroke M
  hpt2 neg 0 R hpt2 vpt2 V stroke } def
/T { stroke [] 0 setdash 2 copy vpt 1.12 mul add M
  hpt neg vpt -1.62 mul V
  hpt 2 mul 0 V
  hpt neg vpt 1.62 mul V closepath stroke
  P  } def
/S { 2 copy A C} def
end
}
\begin{picture}(3600,2160)(0,0)
\special{"
gnudict begin
gsave
50 50 translate
0.100 0.100 scale
0 setgray
/Helvetica findfont 100 scalefont setfont
newpath
-500.000000 -500.000000 translate
LTa
LTb
600 2109 M
63 0 V
2754 0 R
-63 0 V
600 1799 M
63 0 V
2754 0 R
-63 0 V
600 1490 M
63 0 V
2754 0 R
-63 0 V
600 1180 M
63 0 V
2754 0 R
-63 0 V
600 870 M
63 0 V
2754 0 R
-63 0 V
600 561 M
63 0 V
2754 0 R
-63 0 V
600 251 M
63 0 V
2754 0 R
-63 0 V
705 251 M
0 63 V
0 1795 R
0 -63 V
1007 251 M
0 63 V
0 1795 R
0 -63 V
1308 251 M
0 63 V
0 1795 R
0 -63 V
1609 251 M
0 63 V
0 1795 R
0 -63 V
1911 251 M
0 63 V
0 1795 R
0 -63 V
2212 251 M
0 63 V
0 1795 R
0 -63 V
2513 251 M
0 63 V
0 1795 R
0 -63 V
2814 251 M
0 63 V
0 1795 R
0 -63 V
3116 251 M
0 63 V
0 1795 R
0 -63 V
3417 251 M
0 63 V
0 1795 R
0 -63 V
600 251 M
2817 0 V
0 1858 V
-2817 0 V
600 251 L
LT0
1669 1025 M
180 0 V
600 2109 M
8 -13 V
30 -49 V
30 -45 V
30 -40 V
30 -37 V
30 -35 V
30 -32 V
30 -30 V
31 -29 V
30 -26 V
30 -26 V
30 -24 V
30 -23 V
30 -22 V
30 -21 V
30 -20 V
31 -20 V
30 -19 V
30 -18 V
30 -17 V
30 -17 V
30 -17 V
30 -16 V
30 -15 V
31 -15 V
30 -14 V
30 -13 V
30 -13 V
30 -12 V
30 -9 V
30 -10 V
31 -11 V
30 -12 V
30 -12 V
30 -13 V
30 -13 V
30 -14 V
30 -14 V
30 -14 V
31 -14 V
30 -15 V
30 -14 V
30 -15 V
30 -15 V
30 -15 V
30 -15 V
30 -15 V
31 -15 V
30 -16 V
30 -15 V
30 -15 V
30 -15 V
30 -15 V
30 -15 V
30 -16 V
31 -15 V
30 -15 V
30 -15 V
30 -15 V
30 -15 V
30 -15 V
30 -15 V
30 -15 V
31 -15 V
30 -15 V
30 -15 V
30 -15 V
30 -14 V
30 -15 V
30 -15 V
31 -14 V
30 -15 V
30 -14 V
30 -15 V
30 -14 V
30 -15 V
30 -14 V
30 -15 V
31 -14 V
30 -14 V
30 -14 V
30 -15 V
30 -14 V
30 -14 V
30 -14 V
30 -14 V
31 -14 V
30 -14 V
30 -14 V
30 -14 V
30 -13 V
30 -14 V
30 -14 V
30 -14 V
LT1
1669 875 M
180 0 V
600 2109 M
8 -9 V
30 -36 V
30 -33 V
30 -30 V
30 -28 V
30 -26 V
30 -25 V
30 -23 V
31 -22 V
30 -21 V
30 -20 V
30 -19 V
30 -18 V
30 -17 V
30 -16 V
30 -16 V
31 -15 V
30 -14 V
30 -12 V
30 -11 V
30 -12 V
30 -12 V
30 -12 V
30 -13 V
31 -12 V
30 -13 V
30 -13 V
30 -13 V
30 -13 V
30 -13 V
30 -13 V
31 -13 V
30 -12 V
30 -13 V
30 -13 V
30 -12 V
30 -12 V
30 -13 V
30 -12 V
31 -12 V
30 -11 V
30 -12 V
30 -12 V
30 -11 V
30 -12 V
30 -11 V
30 -11 V
31 -11 V
30 -11 V
30 -11 V
30 -11 V
30 -11 V
30 -10 V
30 -11 V
30 -10 V
31 -11 V
30 -10 V
30 -10 V
30 -10 V
30 -10 V
30 -10 V
30 -10 V
30 -9 V
31 -10 V
30 -10 V
30 -9 V
30 -10 V
30 -9 V
30 -9 V
30 -10 V
31 -9 V
30 -9 V
30 -9 V
30 -9 V
30 -9 V
30 -8 V
30 -9 V
30 -9 V
31 -9 V
30 -8 V
30 -9 V
30 -8 V
30 -9 V
30 -8 V
30 -8 V
30 -9 V
31 -8 V
30 -8 V
30 -8 V
30 -8 V
30 -8 V
30 -8 V
30 -8 V
30 -8 V
LT2
1669 725 M
180 0 V
600 2109 M
8 -12 V
30 -44 V
30 -41 V
30 -37 V
30 -35 V
30 -32 V
30 -31 V
30 -28 V
31 -27 V
30 -26 V
30 -24 V
30 -24 V
30 -22 V
30 -21 V
30 -20 V
30 -20 V
31 -19 V
30 -18 V
30 -17 V
30 -17 V
30 -16 V
30 -15 V
30 -14 V
30 -13 V
31 -12 V
30 -12 V
30 -12 V
30 -13 V
30 -13 V
30 -14 V
30 -14 V
31 -14 V
30 -14 V
30 -15 V
30 -14 V
30 -15 V
30 -14 V
30 -15 V
30 -14 V
31 -15 V
30 -14 V
30 -15 V
30 -14 V
30 -14 V
30 -15 V
30 -14 V
30 -14 V
31 -14 V
30 -14 V
30 -14 V
30 -13 V
30 -14 V
30 -14 V
30 -13 V
30 -14 V
31 -13 V
30 -14 V
30 -13 V
30 -13 V
30 -13 V
30 -13 V
30 -13 V
30 -13 V
31 -12 V
30 -13 V
30 -13 V
30 -12 V
30 -13 V
30 -12 V
30 -13 V
31 -12 V
30 -12 V
30 -12 V
30 -12 V
30 -12 V
30 -12 V
30 -12 V
30 -12 V
31 -12 V
30 -11 V
30 -12 V
30 -12 V
30 -11 V
30 -12 V
30 -11 V
30 -11 V
31 -12 V
30 -11 V
30 -11 V
30 -11 V
30 -11 V
30 -12 V
30 -11 V
30 -10 V
LT3
1669 575 M
180 0 V
600 2109 M
8 -13 V
30 -48 V
30 -42 V
30 -39 V
30 -34 V
30 -32 V
30 -28 V
30 -27 V
31 -25 V
30 -23 V
30 -22 V
30 -20 V
30 -19 V
30 -19 V
30 -17 V
30 -17 V
31 -16 V
30 -16 V
30 -15 V
30 -14 V
30 -14 V
30 -14 V
30 -13 V
30 -13 V
31 -12 V
30 -12 V
30 -12 V
30 -11 V
30 -11 V
30 -11 V
30 -10 V
31 -9 V
30 -9 V
30 -8 V
30 -6 V
30 -6 V
30 -8 V
30 -8 V
30 -10 V
31 -9 V
30 -11 V
30 -11 V
30 -12 V
30 -12 V
30 -12 V
30 -13 V
30 -13 V
31 -13 V
30 -14 V
30 -14 V
30 -14 V
30 -14 V
30 -14 V
30 -15 V
30 -14 V
31 -15 V
30 -15 V
30 -15 V
30 -15 V
30 -15 V
30 -15 V
30 -15 V
30 -16 V
31 -15 V
30 -15 V
30 -16 V
30 -15 V
30 -16 V
30 -16 V
30 -15 V
31 -16 V
30 -15 V
30 -16 V
30 -16 V
30 -15 V
30 -16 V
30 -16 V
30 -16 V
31 -15 V
30 -16 V
30 -16 V
30 -15 V
30 -16 V
30 -16 V
30 -16 V
30 -15 V
31 -16 V
30 -16 V
30 -16 V
30 -16 V
30 -15 V
30 -16 V
30 -16 V
30 -15 V
LT4
1669 425 M
180 0 V
600 2109 M
8 -11 V
30 -39 V
30 -34 V
30 -29 V
30 -25 V
30 -21 V
30 -19 V
30 -17 V
31 -15 V
30 -13 V
30 -12 V
30 -11 V
30 -10 V
30 -9 V
30 -8 V
30 -8 V
31 -7 V
30 -7 V
30 -7 V
30 -6 V
30 -6 V
30 -6 V
30 -5 V
30 -6 V
31 -5 V
30 -5 V
30 -5 V
30 -5 V
30 -5 V
30 -5 V
30 -4 V
31 -5 V
30 -4 V
30 -5 V
30 -4 V
30 -3 V
30 -4 V
30 -3 V
30 -1 V
31 0 V
30 0 V
30 -2 V
30 -3 V
30 -4 V
30 -5 V
30 -6 V
30 -6 V
31 -8 V
30 -7 V
30 -9 V
30 -9 V
30 -9 V
30 -10 V
30 -11 V
30 -10 V
31 -11 V
30 -12 V
30 -12 V
30 -12 V
30 -12 V
30 -12 V
30 -13 V
30 -13 V
31 -14 V
30 -13 V
30 -14 V
30 -14 V
30 -14 V
30 -14 V
30 -14 V
31 -14 V
30 -15 V
30 -15 V
30 -15 V
30 -15 V
30 -15 V
30 -15 V
30 -15 V
31 -15 V
30 -16 V
30 -15 V
30 -16 V
30 -16 V
30 -16 V
30 -16 V
30 -16 V
31 -16 V
30 -16 V
30 -16 V
30 -16 V
30 -16 V
30 -17 V
30 -16 V
30 -17 V
stroke
grestore
end
showpage
}
\put(1609,425){\makebox(0,0)[r]{$\Mt = 225 \GeV$}}
\put(1609,575){\makebox(0,0)[r]{$\Mt = 200 \GeV$}}
\put(1609,725){\makebox(0,0)[r]{$\Mt = 150 \GeV$}}
\put(1609,875){\makebox(0,0)[r]{$\Mt = 125 \GeV$}}
\put(1609,1025){\makebox(0,0)[r]{$\Mt = 175 \GeV$}}
\put(3688,151){\makebox(0,0){$\frac{\MH}{\GeV}$}}
\put(40,1080){%
\makebox(0,0)[b]{\shortstack{$\Delta r_{(2),\rm subtr}^{\rm top}$}}%
}
\put(3417,151){\makebox(0,0){1000}}
\put(3116,151){\makebox(0,0){900}}
\put(2814,151){\makebox(0,0){800}}
\put(2513,151){\makebox(0,0){700}}
\put(2212,151){\makebox(0,0){600}}
\put(1911,151){\makebox(0,0){500}}
\put(1609,151){\makebox(0,0){400}}
\put(1308,151){\makebox(0,0){300}}
\put(1007,151){\makebox(0,0){200}}
\put(705,151){\makebox(0,0){100}}
\put(540,251){\makebox(0,0)[r]{$-0.0012$}}
\put(540,561){\makebox(0,0)[r]{$-0.0010$}}
\put(540,870){\makebox(0,0)[r]{$-0.0008$}}
\put(540,1180){\makebox(0,0)[r]{$-0.0006$}}
\put(540,1490){\makebox(0,0)[r]{$-0.0004$}}
\put(540,1799){\makebox(0,0)[r]{$-0.0002$}}
\put(540,2109){\makebox(0,0)[r]{$0$}}
\end{picture}
\ece
\caption{
Two-loop top-quark contribution to
$\Delta r$ subtracted at $\MH=65\,\GeV$.
\label{fig:delr1}}
\end{figure}
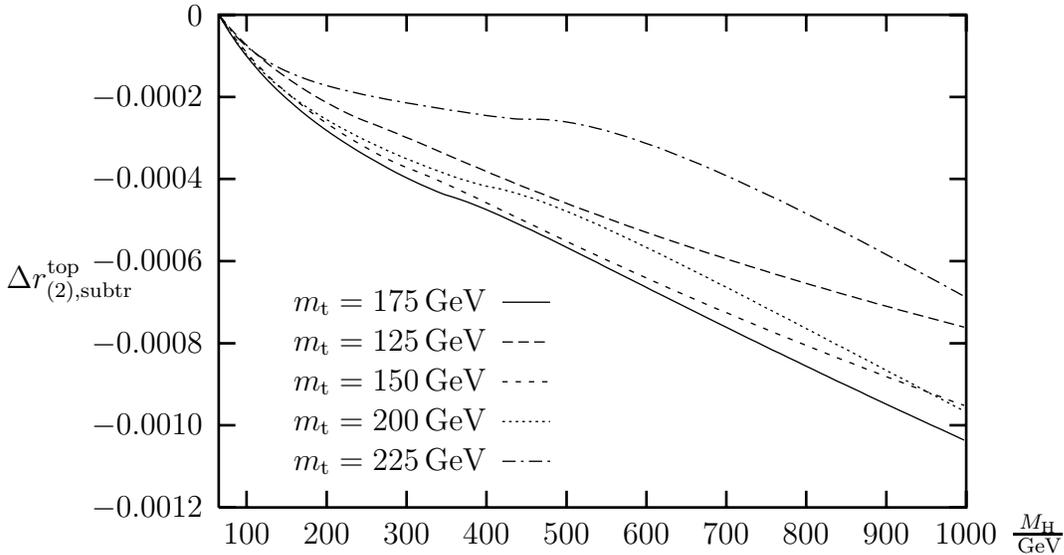

The result for $\De r^{\mathrm top}_{(2), {\mathrm subtr}}(\MH)$
is shown in \reffi{fig:delr1} for various values of $\Mt$. The
Higgs-boson mass is varied in the interval 
$65 \GeV \leq \MH \leq 1 \TeV$. The change in 
$\De r^{\mathrm top}_{(2), {\mathrm subtr}}(\MH)$ over this interval
is about 0.001, which corresponds to a shift in $\MW$ of about 
$20 \MeV$.
It is interesting to note that the absolute value of
the correction is maximal just in the region of $\Mt = 175 \GeV$,
i.e.\ for the physical value of the top-quark mass. 
For $\Mt \sim 175 \GeV$ the correction 
$\De r^{\mathrm top}_{(2), {\mathrm subtr}}(\MH)$
amounts to about $10 \%$ of the one-loop contribution,
$\De r_{(1), {\mathrm subtr}}(\MH)$,
which is defined in analogy to~\refeq{eq:DeltaRsubtr}.
\smallskip

The other $\MH$-dependent two-loop correction that is expected to
be sizable is 
the contribution of the terms proportional to $\De \al$. It reads
\beqar
\De r^{\De\al}_{(2), {\mathrm subtr}}(\MH) & = &
2 \De\al \left[
\frac{\Sigma^{\PW}_{(1), \PH}(0) -
\Re \Sigma^{\PW}_{(1), \PH}(\MW^2)}{\MW^2} - 
2 \frac {\de \swHone}{\sw} \right]_{\mathrm subtr} \nn \\
&=& 2 \De\al \, \De r_{(1), {\mathrm subtr}}(\MH),
\label{eq:DelRbarDelAlp}
\eeqar
and can easily be obtained by a proper resummation of one-loop
terms~\cite{sirresum}. 

The remaining fermionic contribution, 
$\De r^{\mathrm lf}_{(2), {\mathrm subtr}}$,
is the one of the light fermions,
i.e.\ of the leptons and of the quark doublets of the first and second
generation,
which is not contained in $\De\al$. Its structure is analogous to
\refeq{eq:Deltrtopsubtr}, but due to the negligible coupling of
the light fermions to the Higgs boson much less diagrams contribute.
The scalar two-loop integrals needed for the light-fermion
contribution can be solved analytically in terms of polylogarithmic
functions. They can be found in \citere{raibas}.

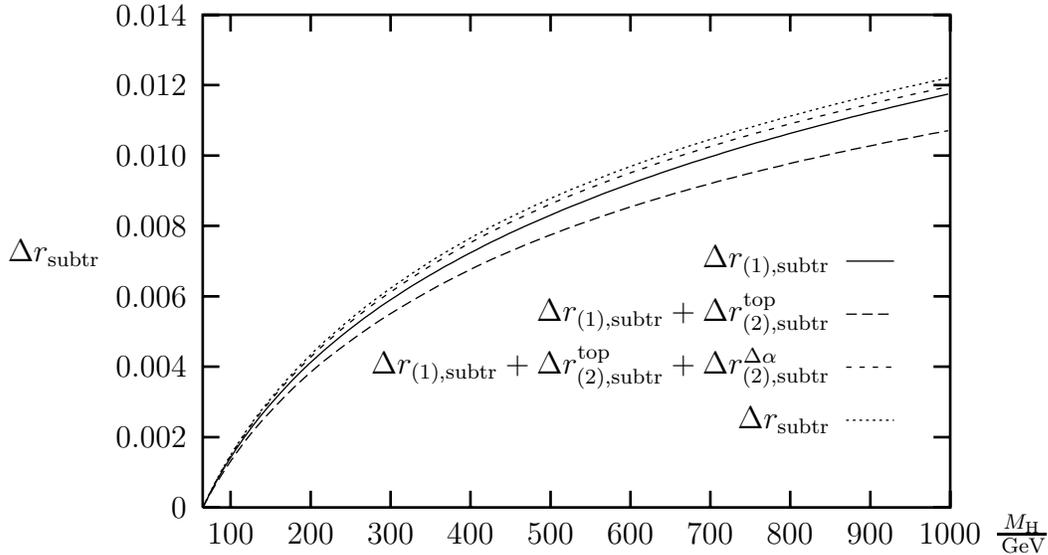
\begin{figure}[htb]
\bce
\setlength{\unitlength}{0.1bp}
\special{!
/gnudict 40 dict def
gnudict begin
/Color false def
/Solid false def
/gnulinewidth 5.000 def
/vshift -33 def
/dl {10 mul} def
/hpt 31.5 def
/vpt 31.5 def
/M {moveto} bind def
/L {lineto} bind def
/R {rmoveto} bind def
/V {rlineto} bind def
/vpt2 vpt 2 mul def
/hpt2 hpt 2 mul def
/Lshow { currentpoint stroke M
  0 vshift R show } def
/Rshow { currentpoint stroke M
  dup stringwidth pop neg vshift R show } def
/Cshow { currentpoint stroke M
  dup stringwidth pop -2 div vshift R show } def
/DL { Color {setrgbcolor Solid {pop []} if 0 setdash }
 {pop pop pop Solid {pop []} if 0 setdash} ifelse } def
/BL { stroke gnulinewidth 2 mul setlinewidth } def
/AL { stroke gnulinewidth 2 div setlinewidth } def
/PL { stroke gnulinewidth setlinewidth } def
/LTb { BL [] 0 0 0 DL } def
/LTa { AL [1 dl 2 dl] 0 setdash 0 0 0 setrgbcolor } def
/LT0 { PL [] 0 1 0 DL } def
/LT1 { PL [4 dl 2 dl] 0 0 1 DL } def
/LT2 { PL [2 dl 3 dl] 1 0 0 DL } def
/LT3 { PL [1 dl 1.5 dl] 1 0 1 DL } def
/LT4 { PL [5 dl 2 dl 1 dl 2 dl] 0 1 1 DL } def
/LT5 { PL [4 dl 3 dl 1 dl 3 dl] 1 1 0 DL } def
/LT6 { PL [2 dl 2 dl 2 dl 4 dl] 0 0 0 DL } def
/LT7 { PL [2 dl 2 dl 2 dl 2 dl 2 dl 4 dl] 1 0.3 0 DL } def
/LT8 { PL [2 dl 2 dl 2 dl 2 dl 2 dl 2 dl 2 dl 4 dl] 0.5 0.5 0.5 DL } def
/P { stroke [] 0 setdash
  currentlinewidth 2 div sub M
  0 currentlinewidth V stroke } def
/D { stroke [] 0 setdash 2 copy vpt add M
  hpt neg vpt neg V hpt vpt neg V
  hpt vpt V hpt neg vpt V closepath stroke
  P } def
/A { stroke [] 0 setdash vpt sub M 0 vpt2 V
  currentpoint stroke M
  hpt neg vpt neg R hpt2 0 V stroke
  } def
/B { stroke [] 0 setdash 2 copy exch hpt sub exch vpt add M
  0 vpt2 neg V hpt2 0 V 0 vpt2 V
  hpt2 neg 0 V closepath stroke
  P } def
/C { stroke [] 0 setdash exch hpt sub exch vpt add M
  hpt2 vpt2 neg V currentpoint stroke M
  hpt2 neg 0 R hpt2 vpt2 V stroke } def
/T { stroke [] 0 setdash 2 copy vpt 1.12 mul add M
  hpt neg vpt -1.62 mul V
  hpt 2 mul 0 V
  hpt neg vpt 1.62 mul V closepath stroke
  P  } def
/S { 2 copy A C} def
end
}
\begin{picture}(3600,2160)(0,0)
\special{"
gnudict begin
gsave
50 50 translate
0.100 0.100 scale
0 setgray
/Helvetica findfont 100 scalefont setfont
newpath
-500.000000 -500.000000 translate
LTa
600 251 M
2817 0 V
LTb
600 251 M
63 0 V
2754 0 R
-63 0 V
600 516 M
63 0 V
2754 0 R
-63 0 V
600 782 M
63 0 V
2754 0 R
-63 0 V
600 1047 M
63 0 V
2754 0 R
-63 0 V
600 1313 M
63 0 V
2754 0 R
-63 0 V
600 1578 M
63 0 V
2754 0 R
-63 0 V
600 1844 M
63 0 V
2754 0 R
-63 0 V
600 2109 M
63 0 V
2754 0 R
-63 0 V
705 251 M
0 63 V
0 1795 R
0 -63 V
1007 251 M
0 63 V
0 1795 R
0 -63 V
1308 251 M
0 63 V
0 1795 R
0 -63 V
1609 251 M
0 63 V
0 1795 R
0 -63 V
1911 251 M
0 63 V
0 1795 R
0 -63 V
2212 251 M
0 63 V
0 1795 R
0 -63 V
2513 251 M
0 63 V
0 1795 R
0 -63 V
2814 251 M
0 63 V
0 1795 R
0 -63 V
3116 251 M
0 63 V
0 1795 R
0 -63 V
3417 251 M
0 63 V
0 1795 R
0 -63 V
600 251 M
2817 0 V
0 1858 V
-2817 0 V
600 251 L
LT0
3025 1180 M
180 0 V
600 251 M
8 15 V
30 59 V
30 54 V
30 49 V
30 46 V
30 43 V
30 41 V
30 38 V
31 37 V
30 34 V
30 33 V
30 32 V
30 30 V
30 29 V
30 27 V
30 27 V
31 26 V
30 25 V
30 24 V
30 23 V
30 23 V
30 21 V
30 22 V
30 20 V
31 20 V
30 20 V
30 19 V
30 18 V
30 18 V
30 17 V
30 18 V
31 16 V
30 17 V
30 16 V
30 15 V
30 16 V
30 15 V
30 14 V
30 15 V
31 14 V
30 14 V
30 13 V
30 13 V
30 13 V
30 13 V
30 13 V
30 12 V
31 12 V
30 12 V
30 12 V
30 12 V
30 11 V
30 11 V
30 11 V
30 11 V
31 11 V
30 10 V
30 11 V
30 10 V
30 10 V
30 10 V
30 10 V
30 9 V
31 10 V
30 9 V
30 10 V
30 9 V
30 9 V
30 9 V
30 9 V
31 8 V
30 9 V
30 8 V
30 9 V
30 8 V
30 8 V
30 8 V
30 8 V
31 8 V
30 8 V
30 8 V
30 8 V
30 7 V
30 8 V
30 7 V
30 7 V
31 8 V
30 7 V
30 7 V
30 7 V
30 7 V
30 7 V
30 7 V
30 7 V
LT1
3025 980 M
180 0 V
600 251 M
8 14 V
30 55 V
30 49 V
30 46 V
30 43 V
30 41 V
30 38 V
30 35 V
31 34 V
30 32 V
30 31 V
30 30 V
30 28 V
30 27 V
30 26 V
30 25 V
31 24 V
30 23 V
30 22 V
30 22 V
30 21 V
30 21 V
30 20 V
30 19 V
31 19 V
30 18 V
30 18 V
30 17 V
30 17 V
30 17 V
30 16 V
31 16 V
30 15 V
30 15 V
30 15 V
30 14 V
30 14 V
30 13 V
30 13 V
31 13 V
30 13 V
30 12 V
30 12 V
30 12 V
30 11 V
30 12 V
30 11 V
31 11 V
30 10 V
30 11 V
30 10 V
30 10 V
30 10 V
30 10 V
30 9 V
31 10 V
30 9 V
30 9 V
30 9 V
30 9 V
30 8 V
30 9 V
30 8 V
31 8 V
30 9 V
30 8 V
30 7 V
30 8 V
30 8 V
30 7 V
31 8 V
30 7 V
30 7 V
30 8 V
30 7 V
30 7 V
30 6 V
30 7 V
31 7 V
30 7 V
30 6 V
30 7 V
30 6 V
30 6 V
30 6 V
30 7 V
31 6 V
30 6 V
30 5 V
30 6 V
30 6 V
30 6 V
30 5 V
30 6 V
LT2
3025 780 M
180 0 V
600 251 M
8 16 V
30 60 V
30 56 V
30 51 V
30 48 V
30 44 V
30 43 V
30 39 V
31 38 V
30 36 V
30 34 V
30 33 V
30 31 V
30 30 V
30 29 V
30 28 V
31 27 V
30 26 V
30 25 V
30 24 V
30 23 V
30 23 V
30 22 V
30 21 V
31 21 V
30 21 V
30 19 V
30 20 V
30 19 V
30 18 V
30 18 V
31 18 V
30 17 V
30 17 V
30 16 V
30 16 V
30 15 V
30 15 V
30 15 V
31 14 V
30 14 V
30 14 V
30 13 V
30 14 V
30 13 V
30 12 V
30 13 V
31 12 V
30 12 V
30 11 V
30 12 V
30 11 V
30 11 V
30 11 V
30 11 V
31 10 V
30 11 V
30 10 V
30 10 V
30 10 V
30 10 V
30 9 V
30 10 V
31 9 V
30 9 V
30 9 V
30 9 V
30 9 V
30 8 V
30 9 V
31 8 V
30 8 V
30 9 V
30 8 V
30 8 V
30 7 V
30 8 V
30 8 V
31 7 V
30 8 V
30 7 V
30 7 V
30 8 V
30 7 V
30 7 V
30 7 V
31 7 V
30 6 V
30 7 V
30 7 V
30 6 V
30 7 V
30 6 V
30 6 V
LT3
3025 580 M
180 0 V
600 251 M
8 16 V
30 61 V
30 57 V
30 52 V
30 48 V
30 46 V
30 43 V
30 40 V
31 38 V
30 37 V
30 35 V
30 33 V
30 32 V
30 30 V
30 30 V
30 28 V
31 27 V
30 27 V
30 25 V
30 25 V
30 24 V
30 23 V
30 22 V
30 22 V
31 22 V
30 20 V
30 21 V
30 19 V
30 20 V
30 19 V
30 18 V
31 18 V
30 17 V
30 17 V
30 17 V
30 16 V
30 16 V
30 15 V
30 15 V
31 15 V
30 14 V
30 15 V
30 13 V
30 14 V
30 13 V
30 13 V
30 13 V
31 12 V
30 12 V
30 12 V
30 12 V
30 12 V
30 11 V
30 11 V
30 11 V
31 11 V
30 11 V
30 10 V
30 11 V
30 10 V
30 10 V
30 10 V
30 9 V
31 10 V
30 9 V
30 9 V
30 10 V
30 9 V
30 8 V
30 9 V
31 9 V
30 8 V
30 9 V
30 8 V
30 8 V
30 8 V
30 8 V
30 8 V
31 8 V
30 8 V
30 7 V
30 8 V
30 7 V
30 7 V
30 8 V
30 7 V
31 7 V
30 7 V
30 7 V
30 6 V
30 7 V
30 7 V
30 6 V
30 7 V
stroke
grestore
end
showpage
}
\put(2965,580){\makebox(0,0)[r]{$\Delta r_{\rm subtr}$}}
\put(2965,780){\makebox(0,0)[r]{$\Delta r_{(1),\rm subtr}+     \Delta r_{(2),\rm subtr}^{{\rm top}}+       \Delta r_{(2),\rm subtr}^{\Delta \alpha }$}}
\put(2965,980){\makebox(0,0)[r]{$\Delta r_{(1),\rm subtr}+ \Delta r_{(2),\rm subtr}^{\rm top}$}}
\put(2965,1180){\makebox(0,0)[r]{$\Delta r_{(1),\rm subtr}$}}
\put(3688,151){\makebox(0,0){$\frac{\MH}{\GeV}$}}
\put(40,1180){%
\makebox(0,0)[b]{\shortstack{$\Delta r_{\rm subtr}$}}%
}
\put(3417,151){\makebox(0,0){1000}}
\put(3116,151){\makebox(0,0){900}}
\put(2814,151){\makebox(0,0){800}}
\put(2513,151){\makebox(0,0){700}}
\put(2212,151){\makebox(0,0){600}}
\put(1911,151){\makebox(0,0){500}}
\put(1609,151){\makebox(0,0){400}}
\put(1308,151){\makebox(0,0){300}}
\put(1007,151){\makebox(0,0){200}}
\put(705,151){\makebox(0,0){100}}
\put(540,2109){\makebox(0,0)[r]{$0.014$}}
\put(540,1844){\makebox(0,0)[r]{$0.012$}}
\put(540,1578){\makebox(0,0)[r]{$0.010$}}
\put(540,1313){\makebox(0,0)[r]{$0.008$}}
\put(540,1047){\makebox(0,0)[r]{$0.006$}}
\put(540,782){\makebox(0,0)[r]{$0.004$}}
\put(540,516){\makebox(0,0)[r]{$0.002$}}
\put(540,251){\makebox(0,0)[r]{$0$}}
\end{picture}
\ece
\caption{
One-loop and two-loop contributions to $\Delta r$ 
subtracted at $\MH=65\,\GeV$.
$\De r_{\mathrm subtr}$ is the result for the full one-loop and fermionic 
two-loop contributions to $\De r$, as defined in the text.}
\label{fig:delr2}
\end{figure}

The total result for the one-loop and fermionic two-loop contributions
to $\De r$, subtracted at $\MH=65\,\GeV$, reads
\beq
\De r_{\mathrm subtr} \equiv \De r_{(1), {\mathrm subtr}} +
\De r^{\mathrm top}_{(2), {\mathrm subtr}} + 
\De r^{\De\al}_{(2), {\mathrm subtr}} + 
\De r^{\mathrm lf}_{(2), {\mathrm subtr}} .
\eeq
It is shown in \reffi{fig:delr2}, where separately also 
the one-loop contribution
$\De r_{(1), {\mathrm subtr}}$, as well as 
$\De r_{(1), {\mathrm subtr}} + \De r^{\mathrm top}_{(2), {\mathrm
subtr}}$, and
$\De r_{(1), {\mathrm subtr}} + \De r^{\mathrm top}_{(2), {\mathrm
subtr}} + \De r^{\De\al}_{(2), {\mathrm subtr}}$
are shown for $\Mt = 175.6 \GeV$.
In \refta{tab:Deltr} numerical values for the different contributions
are given for several values of $\MH$.
It can be seen that the higher-order contributions 
$\De r^{\mathrm top}_{(2), {\mathrm subtr}}(\MH)$ and 
$\De r^{\De\al}_{(2), {\mathrm subtr}}(\MH)$ are of about the same size
and to a large extent cancel each other. 
The light-fermion contributions which are not contained in $\De\al$
add a relatively small correction. Over the full range of the
Higgs-boson mass it amounts to about $4\, \MeV$.
In total, the inclusion of the higher-order contributions discussed here
leads to a slight increase in the sensitivity to the Higgs-boson mass 
compared to the pure one-loop result.

\btab
$$
\barr{|c||c|c|c|c|} \hline
\MH/\GeV & \De r_{(1), {\mathrm subtr}}/10^{-3} & 
	   \De r^{\mathrm top}_{(2), {\mathrm subtr}}/10^{-3} & 
           \De r^{\De\al}_{(2), {\mathrm subtr}}/10^{-3} & 
           \De r_{\mathrm subtr}/10^{-3} \\ \hline \hline
65  & 0 & 0 & 0 &  0 \\ \hline
100 & 1.42 & -0.101 & 0.148 & 1.49 \\ \hline
200 & 4.12 & -0.282 & 0.431 & 4.34 \\ \hline
300 & 5.90 & -0.397 & 0.620 & 6.23 \\ \hline
400 & 7.24 & -0.474 & 0.762 & 7.66 \\ \hline
500 & 8.31 & -0.565 & 0.876 & 8.78 \\ \hline
600 & 9.20 & -0.663 & 0.971 & 9.69 \\ \hline
1000 & 11.8\phantom{00} & -1.04\phantom{0} & 1.25\phantom{0} & 
12.2\phantom{00} \\ \hline
\earr
$$
\caption{The dependence of one-loop and two-loop contributions to 
$\De r$ on the Higgs-boson mass for $\Mt = 175.6\, \GeV$ (see text).
\label{tab:Deltr}}
\etab

Regarding the remaining Higgs-mass dependence of $\De r$ at the
two-loop level, there are only purely bosonic corrections left, which
contain no specific source of enhancement. They
can be expected to yield a contribution to 
$\De r_{(2), {\mathrm subtr}}(\MH)$ of about the same size as
$\left.\left(\De r^{\mathrm bos}_{(1)}(\MH)\right)^2\right|_{\mathrm
subtr}$, where $\De r^{\mathrm bos}_{(1)}$ denotes the bosonic
contribution to $\De r$ at the one-loop level. The contribution of
$\left.\left(\De r^{\mathrm bos}_{(1)}(\MH)\right)^2\right|_{\mathrm
subtr}$ amounts to only about $10 \%$ of $\De r^{\mathrm top}_{(2), 
{\mathrm subtr}}(\MH)$ corresponding to a shift of about $2 \MeV$ in the
W-boson mass. 
This estimate agrees well with the values obtained for the 
Higgs-mass dependence from the formula in the second paper of
\citere{qcd2} for the leading term proportional to $\MH^2$ in an 
asymptotic expansion for large Higgs-boson mass. 
The Higgs-mass dependence of the term proportional to $\MH^2$ amounts to 
less than $15 \%$ of $\De r^{\mathrm top}_{(2), {\mathrm subtr}}(\MH)$ for
reasonable values of $\MH$.

\smallskip

The result for 
$\De r_{\mathrm subtr}^{{\mathrm top}, \De\al} \equiv
\De r_{(1), {\mathrm subtr}} + \De r^{\mathrm top}_{(2), {\mathrm
subtr}} + \De r^{\De\al}_{(2), {\mathrm subtr}}$
can be compared to the result
obtained via an expansion in $\Mt$ up
to next-to-leading order, i.e.\
${\cal O}(\GF^2 \Mt^2 \MZ^2)$ \cite{gamb,DGS}.
The results for $\MW$ as a function of $\MH$ according to this
expansion (without QCD corrections; $\Mt = 175.6\, \GeV$) are given
in \refta{tab:MWgamb}~\cite{gambpriv}.
Extracting from \refta{tab:MWgamb} the corresponding values of 
$\De r$ and subtracting at $\MH = 65\, \GeV$ yields the values 
$\De r_{\mathrm subtr}^{{\mathrm top}, \De\al, {\mathrm expa}}(\MH)$ 
as results of the expansion in $\Mt$. These are compared to 
the exact result
$\De r_{\mathrm subtr}^{{\mathrm top}, \De\al}(\MH)$ in 
\refta{tab:MWcomp}. In the last column of \refta{tab:MWcomp} the 
approximate shift in $\MW$ is given which corresponds to the difference 
between exact result and expansion.
The results agree within about $30 \%$ of 
$\De r^{\mathrm top}_{(2), {\mathrm subtr}}(\MH)$,
which amounts to a difference in $\MW$ of up to about $4\, \MeV$.

\btab
$$
\barr{|c||c|c|c|c|c|} \hline
\MH/\GeV & 65 & 100 & 300 & 600 & 1000\\ \hline
\MW/\GeV & 80.4819 & 80.4584 & 80.3837 & 80.3294 & 80.2901 \\ \hline
\earr
$$
\caption{The results for $\MW$ as a function of $\MH$ (without QCD
corrections; $\Mt = 175.6\, \GeV$) obtained via an expansion up to
next-to-leading order in $\Mt$~\cite{gambpriv}.
\label{tab:MWgamb}}
\etab

\btab
$$
\barr{|c||c|c||c|} \hline
\MH/\GeV & 
\De r_{\mathrm subtr}^{{\mathrm top}, \De\al}/10^{-3} &
\De r^{{\mathrm top}, \De\al, {\mathrm expa}}_{\mathrm subtr}/10^{-3} & 
\de \MW/\MeV\\ \hline
65   & 0    & 0    & 0   \\ \hline
100  & 1.48 & 1.52 & 0.6 \\ \hline
300  & 6.16 & 6.32 & 2.5 \\ \hline
600  & 9.56 & 9.79 & 3.6 \\ \hline
1000 & 12.0\phantom{00} & 12.3\phantom{00} & 4.1 \\ \hline
\earr
$$
\caption{Comparison between the exact result, 
$\De r_{\mathrm subtr}^{{\mathrm top}, \De\al}(\MH)$, 
and the result of an expansion up to next-to-leading order in $\Mt$,
$\De r^{{\mathrm top}, \De\al, {\mathrm expa}}_{\mathrm
subtr}(\MH)$. In the last column the approximate shift
in $\MW$ is displayed 
which corresponds to the difference between the two results.
\label{tab:MWcomp}}
\etab

In \refta{tab:DeltaMW} the shift in $\MW$ corresponding to
$\De r_{\mathrm subtr}(\MH)$, i.e.~the change in the theoretical
prediction for $\MW$ when varying the Higgs-boson mass from
$65\, \GeV$ to $1\, \TeV$, is shown for three values of the
top-quark mass, $\Mt = 170, 175, 180\, \GeV$. The dependence
on the precise value of $\Mt$ is rather mild, which is expected 
from the fact that $\Mt$ enters here only at the two-loop level and
that $\De r^{\mathrm top}_{(2), {\mathrm subtr}}(\MH)$ has a local
maximum in the region of $\Mt = 175\, \GeV$ (see \reffi{fig:delr1}).
{}From the shift in $\MW$ given in \refta{tab:DeltaMW} the theoretical
prediction for the absolute value of $\MW$ can be obtained using 
as input one value of $\MW$ for a given $\MH$. From \citere{DGPS} 
we infer for the subtraction
point $\MH = 65\, \GeV$ the corresponding values 
$\MW = 80.374, 80.404, 80.435\, \GeV$ for
$\Mt = 170, 175, 180\, \GeV$,
respectively.
The accuracy of this subtraction point, being taken from the expansion
up to ${\cal O}(\GF^2 \Mt^2 \MZ^2)$, is of course lower than the
accuracy of the shift in $\MW$ as given in \refta{tab:DeltaMW}.

\btab
$$
\barr{|c||c|c|c|c|c|c|c|c|} \hline
\MH/\GeV & 65 & 100 & 200 & 300 & 400 & 500 & 600 & 1000 \\ \hline
\hline
\Delta \MW(\MH)/\MeV,\; \Mt=170\,\GeV &
0 & -22.6 & -65.8 & -94.5 & -116 & -133 & -147 & -185  \\ \hline
\Delta \MW(\MH)/\MeV,\; \Mt=175\,\GeV &
0 & -22.8 & -66.3 & -95.2 & -117 & -134 & -148 & -187  \\ \hline
\Delta \MW(\MH)/ \MeV,\; \Mt=180\,\GeV &
0 & -23.0 & -66.8 & -96.0 & -118 & -135 & -149 & -188  \\ \hline
\earr
$$
\caption{The shift in the 
theoretical prediction for $\MW$ caused by varying the Higgs-boson mass 
in the interval $65\, \GeV \leq \MH \leq 1\, \TeV$
for three values of $\Mt$. 
\label{tab:DeltaMW}}
\etab

\bigskip

In summary, we have discussed the Higgs-mass dependence of the
two-loop corrections to $\De r$ by considering the subtracted 
quantity $\De r_{\mathrm subtr}(\MH) = \De r(\MH) - 
\De r(\MH = 65\, \GeV)$. Exact results have been presented for
the Higgs-dependent fermionic contributions, i.e.\ no expansion in the 
top-quark and the Higgs-boson mass has been made.
The contribution associated with the top quark has been
compared with the result of an expansion up to next-to-leading order in
$\Mt$. Agreement within about $30 \%$ of the two-loop top-quark
correction has
been found, which corresponds to a difference in $\MW$ of 
about $4\, \MeV$ in the range $65\, \GeV \leq \MH \leq 1\, \TeV$ of the
Higgs-boson mass. 
The Higgs-dependence of the light-fermion contributions leads to a shift 
of $\MW$ of up to $4\, \MeV$. 
The only missing part in the Higgs-dependence of $\De r$
at the two-loop level
are the purely bosonic contributions, which have been estimated to
yield a relatively small correction of up to 
about $2\, \MeV$ in the W-boson mass.
Considering the envisaged 
experimental error of $\MW$ from the measurements at LEP2 and the
Tevatron of $\sim 20 \MeV$, we conclude that
the theoretical uncertainties due to
unknown higher-order corrections in the Higgs-mass dependence of
$\De r$ are now under control.
\bigskip

\vspace{- .3 cm}
\section*{Acknowledgements}

We thank M.~B\"ohm, P.~Gambino, W.~Hollik and A.~Stremplat for useful
discussions. We also thank P.~Gambino for sending us the results
displayed in \refta{tab:MWgamb}.

\end{document}